\documentclass{article}

\usepackage{microtype}
\usepackage{graphicx}
\usepackage{subcaption}
\usepackage{booktabs} 

\usepackage{tabularx}
\usepackage{threeparttable}

\usepackage{hyperref}

\usepackage[preprint]{neurips_2026}

\usepackage{amssymb}
\usepackage{mathtools}
\usepackage{amsthm}
\usepackage{booktabs}
\usepackage{siunitx}
\usepackage{makecell}
\usepackage{graphicx}
\usepackage{array}
\usepackage{algorithm}
\usepackage{algorithmic}
\usepackage{booktabs}
\usepackage{enumitem}
\usepackage{graphicx}
\usepackage{color}
\usepackage{multirow}
\usepackage{colortbl}
\usepackage{xspace}
\usepackage{booktabs}
\usepackage{xcolor}
\usepackage{amssymb} 
\usepackage{pifont}  
\usepackage{tcolorbox}
\usepackage{graphicx}
\usepackage{url} 
\usepackage{tabularx}
\usepackage{makecell}

\usepackage[capitalize,noabbrev]{cleveref}

\theoremstyle{plain}

\theoremstyle{definition}

\theoremstyle{remark}

\newcommand{\srup}[1]{\makebox[2em][l]{\,{\textcolor{ForestGreen}{\scriptsize$\uparrow$#1}}}}

\renewcommand{\srup}[1]{\textsuperscript{\,$\uparrow$#1}}
\newcommand{\srdown}[1]{\textsuperscript{\,$\downarrow$#1}}

\usepackage[textsize=tiny]{todonotes}

\title{OASIF: An Efficient Obfuscation-Aware Self-Improving Framework for LLM-Based Assembly Code Instruction Following and Comprehension}

\author{%
  Xinyi Wang\textsuperscript{\normalfont{1,$\ast$}},
  Rongze Chen\textsuperscript{\normalfont{1}},
  Ke Wang\textsuperscript{\normalfont{2}},
  Qiyuan Chen\textsuperscript{\normalfont{3}},
  Yanming Liu\textsuperscript{\normalfont{4}},
  Xiang Li\textsuperscript{\normalfont{1}},
  Chunfu Jia\textsuperscript{\normalfont{1}
  }
  \\
  \textsuperscript{1} College of Cryptology and Cyber Science, Nankai University \\
  \textsuperscript{2} Renmin University of China \\
  \textsuperscript{3} Ant Group \\
  \textsuperscript{4} Zhejiang University \\
  \texttt{\{2120230741@mail.nankai.edu.cn\}} \\
}

\begin{document}

\maketitle

\begin{abstract}
Large Language Models (LLMs) have recently shown promise in automated binary analysis, yet they remain brittle under commercial-grade obfuscation. We present OASIF, an Obfuscation-Aware Self-evolving Instruction-Following framework for obfuscated assembly comprehension. OASIF couples a token-efficient assembly encoder with a lightweight projector to expose long obfuscated code to a pretrained code LLM under a bounded context budget and follows a three-phase training: (i) feature-space alignment, (ii) supervised instruction fine-tuning, and (iii) online self-evolving reinforcement learning with hybrid rewards, enabling continual adaptation with minimal manual verification. 
On VMISA-Bench, a challenging out-of-distribution suite featuring three commercial VM-based obfuscators, OASIF consistently improves open-source backbones; Qwen2.5-Coder-Instruct-14B attains Success Rate gains of $+15.9$, $+5.8$, and $+16.9$ percentage points (pp) on Code Virtualizer, Themida (v3.0.7), and VMProtect (v3.5), respectively, and improves the OASIF-Bench average by $+9.8$. OASIF further delivers stable gains across seven standard BCSD benchmarks while preserving general and domain-relevant capabilities on HumanEval, VulBench, and HumanEval-Decompile.
\end{abstract}

\vspace{-2mm}

\section{Introduction}

The analysis and comprehension of assembly code underpin a wide range of security-critical tasks, including reverse engineering~\citep{kargen2017towards, megira2018malware}, malware analysis~\citep{megira2018malware}, vulnerability detection~\citep{mantovani2022convergence, taviss2024asm2seq}, and software hardening~\citep{thirumoorthy2022feature}. Unlike high-level programming languages, assembly code exposes low-level hardware operations with minimal syntactic structure and weak semantic abstraction, resulting in long instruction sequences with low information density~\citep{wang2022jtrans} that are difficult to interpret and reason about. Bridging such hardware-centric representations with natural language semantics is therefore essential for downstream tasks such as semantic recovery, behavioral explanation~\citep{tan2024llm4decompile}, and deobfuscation~\citep{al2023extending}, yet it remains a long-standing challenge.

Recent years have witnessed growing interest in applying LLMs~\citep{achiam2023gpt, touvron2023llama, bai2023qwen} to binary analysis~\citep{Microsoft-2025-05-06}. Early efforts predominantly rely on masked language modeling or contrastive representation learning to extract semantic embeddings from assembly code, exemplified by Asm2Vec~\citep{ding2019asm2vec}, CodeArt~\citep{su2024codeart}, and CLAP~\citep{wang2024clap}. These methods capture instruction-level semantics and support similarity-based retrieval, but they are not designed for instruction following or multi-step reasoning. More recent approaches adapt decoder-based LLMs to assembly code through structural modeling or supervised fine-tuning~\citep{jiang2023nova, tan2024llm4decompile, wang2025asma}, achieving improved generation and explanation capabilities. However, most existing methods depend on static supervision, making them brittle under heavy obfuscation and difficult to scale to real-world scenarios.

This limitation becomes particularly pronounced under industrial-grade code obfuscation. Among various techniques, virtual machine (VM)-based obfuscation is widely regarded as one of the strongest protections used in commercial software and real-world malware~\citep{banescu2016code, blazytko2017syntia}. By translating native instructions into proprietary virtual instruction sets executed by custom interpreters, VM-based obfuscators fundamentally reshape control flow, instruction semantics, and execution context~\citep{salwan2018symbolic, li2022chosen}. The resulting binaries exhibit long execution traces, frequent decoy context switches, and complex one-to-many instruction mappings that are absent from standard training corpora~\citep{coogan2011deobfuscation, xu2018vmhunt}. 

Two key challenges hinder LLMs' effectiveness: obfuscated code often expands to hundreds of thousands of instructions, far exceeding the context window of most open-source LLMs, and acquiring fine-grained supervision is prohibitively expensive, especially for commercial black-box obfuscators.

\begin{figure*}[h]
    \centering
    \vspace{-6mm}
    \includegraphics[width=1.0\linewidth]{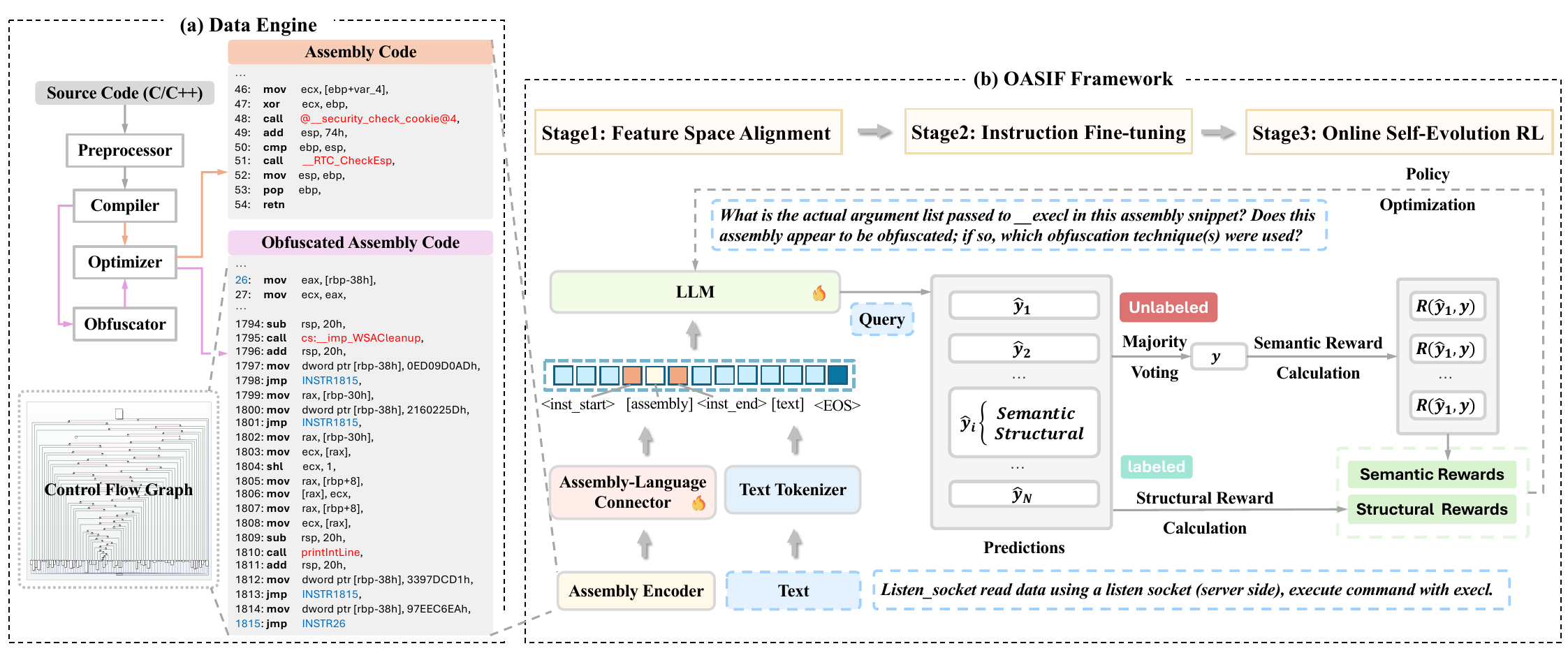}
    \vspace{-6mm}
    \caption{\textbf{Overview of the OASIF framework.} \textbf{(a) Data Engine} generates paired normal/obfuscated assembly by compiling source code and applying obfuscation. \textbf{(b) OASIF Framework} aligns assembly features with a pretrained code LLM and trains in three stages.}
    \vspace{-4mm}
    \label{fig:framework}
\end{figure*}

To address these challenges, we propose OASIF, an Obfuscation-Aware Self-evolving Instruction-Following framework for understanding obfuscated assembly code. As illustrated in Fig.~\ref{fig:framework}, OASIF integrates a token efficient assembly encoder with a pretrained code LLM through a lightweight projection module. We introduce a three-phase training, the first two phases establish stable assembly–language alignment and instruction following behavior via Feature Space Alignment and Instruction Fine-tuning. In the Online Self-Evolving Reinforcement Learning (RL) phase, evaluating its own outputs with hybrid structural and semantic rewards, OASIF enables continuous self-improvement on challenging obfuscated samples.

Our main contributions are summarized as follows.
\begin{itemize}
\vspace{-2mm}
\item We propose OASIF, a framework that introduces a three-phase training featuring online self-evolving reinforcement learning to improve instruction following and comprehension under assembly obfuscation.

\vspace{-2mm}
\item We develop an instruction-centric synthetic data generation engine that enables annotation-efficient adaptation to obfuscated assembly code while maintaining training scalability. Additionally, we present OASIF-Bench, a comprehensive obfuscated assembly comprehension and instruction following benchmark.

\vspace{-2mm}
\item On VMISA-Bench featuring three commercial VM obfuscators, OASIF consistently improves open-source backbones; specifically, Qwen2.5-Coder-Instruct-14B achieves Success Rate (SR) gains of $+15.9$, $+5.8$, and $+16.9$ pp on Code Virtualizer, Themida(v3.0.7), and VMProtect(v3.5), respectively, and improves the OASIF-Bench average score by $+9.8$. Furthermore, OASIF enhances general assembly comprehension across seven standard BCSD benchmarks while preserving domain-relevant capabilities on HumanEval, VulBench, and HumanEval-Decompile.

\end{itemize}

\section{Related Work}

\vspace{-2mm}
\subsection{LLMs for Binary Analysis and Obfuscation}
\vspace{-2mm}

Deep learning for binary analysis has progressed from static representation learning to LLM-enabled generative reasoning. Early work mainly adopts Masked Language Modeling (MLM)~\citep{devlin2019bert} to capture instruction semantics. Asm2Vec~\citep{ding2019asm2vec} and CodeArt~\citep{su2024codeart} learn robust assembly embeddings, and CLAP~\citep{wang2024clap} aligns assembly with natural language via contrastive learning. 

To enhance comprehension, Nova~\citep{jiang2023nova} models assembly structures with hierarchical attention, and ASMA-Tune~\citep{wang2025asma} further aligns assembly features with LLMs through structural semantic tuning. These methods strengthen code LLMs on low-information-density assembly inputs, yet robustness under commercial-grade obfuscation remains challenging.

Obfuscation further increases difficulty due to substantial control-flow and instruction-level perturbations. DisasLLM~\citep{rong2024disassembling} explores an LLM-driven strategy for analyzing executables. However, systematic evaluations~\citep{tkachenko2025deconstructing} show that although models can handle some common transformations, stronger techniques remain difficult, underscoring the need for more robust reasoning and adaptation. Meanwhile, the prevailing reliance on supervised fine-tuning is increasingly limited by the scarcity and cost of high-quality labeled data for obfuscated binaries.

\vspace{-2mm}
\subsection{RL and Self-Evolution in LLMs}
\vspace{-2mm}

Recent progress in self-improvement enables models to enhance inference performance with minimal or no additional annotation~\citep{wang2023self, luo2023wizardcoder, yuan2024self}. Self-consistency~\citep{zuo2025ttrl}, for instance, uses majority voting and consistency signals to stabilize reasoning trajectories. While such paradigms have been shown to be effective in domains like mathematical reasoning and code generation~\citep{guo2025deepseek}, their practical instantiation for binary security tasks remains underexplored, especially where supervision is expensive and data distributions shift sharply under obfuscation.

Our work targets this gap by introducing OASIF, a framework that  integrates a multi-modal assembly encoder with a self-evolving reinforcement learning strategy to enhance instruction understanding and reasoning under obfuscation. Moving beyond purely static supervision, OASIF leverages unlabeled or weakly supervised data to iteratively refine instruction-following and comprehension under obfuscation while also improving generalization on standard assembly understanding tasks.
\vspace{-2mm}
\section{Dataset Engine}\label{sec:dataset_engine}
\vspace{-2mm}

\subsection{Binary Dataset Construction}\label{sec:binary_dataset_construction}
\vspace{-2mm}

\smallskip

\noindent\textbf{Raw corpora.}
Following CLAP~\citep{wang2024clap} and ASMA-Tune~\citep{wang2025asma}, we build a large-scale multi-source corpus from BinaryCorp-3M~\citep{wang2022jtrans} and the Juliet Test Suite~\citep{boland2012juliet}, and further augment Juliet with obfuscated variants to better reflect real-world reverse-engineering settings. BinaryCorp-3M, collected from Arch Linux official repositories and AUR~\citep{Arch-Linux-Package-Search,AUR-(en)-Home}, contains 10{,}265 binaries and $\sim$3M functions across diverse software categories, supports \texttt{gcc}/\texttt{clang} and five optimization levels, and is among the largest corpora for binary code similarity detection (BCSD). Juliet provides 64{,}099 vulnerability-analysis test cases with accompanying vulnerability descriptions, and its semantic annotations have been validated for code understanding tasks~\citep{taviss2024asm2seq}.

\vspace{-2mm}
\smallskip
\noindent\textbf{Compilation, disassembly, and annotation.}

For the BinaryCorp-3M training split, we parse instructions and CFGs with Capstone, reconstruct per-instruction jump metadata, sample across compilers and optimization levels, and retain functions with 10–3,000 instructions. For Juliet, we compile C/C++ sources with \texttt{gcc} under randomly selected optimization levels (\texttt{-O0}--\texttt{-O3}), disassemble the binaries with IDA Pro, extract function names and types from symbol-preserved binaries via IDA’s function identification, and derive function-level textual descriptions from vulnerability headers.

\vspace{-2mm}
\smallskip
\noindent\textbf{Obfuscated variants.}
We further compile Juliet with Obfuscator-LLVM (OLLVM)~\citep{junod2015obfuscator} using instruction substitution (SUB), bogus control flow (BCF), and control-flow flattening (FLA). We disassemble x86\_64 binaries with IDA Pro and rebase instruction addresses while keeping relative jump offsets~\citep{wang2024clap, wang2025asma} to preserve structural semantics and critical identifiers (e.g., function calls and variable names). The resulting raw assembly corpus contains 513{,}106 snippets: 212{,}117 from the BinaryCorp-3M training split, 79{,}920 from unobfuscated Juliet, and 221{,}069 from OLLVM-obfuscated Juliet, including 2{,}651 SUB, 58{,}578 FLA, 79{,}920 BCF, and 79{,}920 ALL (SUB+FLA+BCF) samples. Details and exact OLLVM flags are reported in Appendix~\ref{appx:dataset_composition}.

\vspace{-2mm}
\subsection{Instruction-Centric Synthetic Data Generation}\label{sec:instruction_centric_data_generation}
\vspace{-2mm}

Given an assembly corpus $C=\{c_i\}_{i=1}^{N}$, we use an LLM $M_{\phi}$ to generate instruction-centric supervision via task-specific prompts. For a task type $t\in T$, the generator is defined as:

\vspace{-4mm}
\begin{equation}
\small
G(c,t)=M_{\phi}\!\left(\pi_t(c)\right)\mapsto (Q_t,A_t)
\label{eq:gen}
\end{equation}

\vspace{-2mm}

where $\pi_t(\cdot)$ denotes a few-shot prompt for task $t$. We consider five task types
$T=\{\texttt{simp},\texttt{detail},\texttt{conv},\texttt{reason},\texttt{reason}^+\}$,
which correspond to simplified description, detailed description, multi-turn conversation, reasoning, and challenge-level reasoning, following the perception-to-reasoning progression~\citep{liu2024visual, chen2024sharegpt4v, luo2023wizardcoder}.

\vspace{-2mm}
\smallskip
\noindent\textbf{Three-phase training datasets.}
We organize all synthetic data into three stage-specific datasets:

\vspace{-6pt}
{
\begin{equation}
\small
D_{\text{align}} = \bigcup_{c\in C_{\text{align}}}\left\{G(c,\texttt{simp}) \oplus c\right\}
\label{eq:dpt}
\end{equation}
}
\vspace{-2pt}
{
\begin{equation}
\small
D_{\text{sft}} = \bigcup_{c\in C_{\text{sft}}}\{G(c,t)\oplus c \mid t\in\{\texttt{detail},\texttt{conv},\texttt{reason}\}\}
\label{eq:dsft}
\end{equation}
}
\vspace{-2pt}
{
\begin{equation}
\small
D_{\text{rl}} = \{(Q,A)\}_{j=1}^{M},\ (Q,A)=G(c,\texttt{reason}^+),~c\in C_{\text{rl}}
\label{eq:drl}
\end{equation}
}
\vspace{-4mm}

where $\oplus$ concatenates $c$ with the generated QA pair, and \texttt{reason}$^{+}$ denotes challenge prompts requiring multi-step semantic inference. The instruction mixture in $D_{\text{sft}}$ follows~\citep{liu2024visual}, whereas $D_{\text{rl}}$ requires complete reasoning steps and manual verification.

\vspace{-2mm}
\smallskip
\noindent
Overall, $D=\{D_{\text{align}},D_{\text{sft}},D_{\text{rl}}\}$ provides multi-granularity supervision for three-phase training, emphasizing obfuscation robustness and reasoning-intensive alignment. After stage-specific filtering and deduplication, $D_{\text{align}}$, $D_{\text{sft}}$, and $D_{\text{rl}}$ contain 430{,}027 \texttt{simp}, 11{,}750 \texttt{detail}/\texttt{conv}/\texttt{reason}, and 4{,}244 manually verified \texttt{reason}$^{+}$ instances, respectively. Details are provided in Appendix~\ref{appx:data_quality_control}.
\vspace{-2mm}
\section{Methodology}
\label{sec:method}
\vspace{-2mm}

We present OASIF, an effective obfuscation-aware framework for improving LLMs in assembly instruction following and comprehension. As shown in Fig.~\ref{fig:framework}, OASIF combines an assembly encoder, a lightweight projector, and a pretrained code LLM with a compact three-phase training pipeline. The training pipeline consists of three phases: feature-space alignment (Phase I), supervised instruction fine-tuning (Phase II), and Online Self-Evolving Reinforcement Learning with hybrid rewards (Phase III). Phases I--II establish stable assembly--language alignment and basic instruction-following ability, while Phase III further enhances the model through self-evolving RL, enabling continual refinement without relying on abundant ground-truth annotations for commercially obfuscated code. 

\vspace{-2mm}
\subsection{Framework Structure}
\label{sec:model_structure}
\vspace{-2mm}

\paragraph{Token-Efficient Assembly Representation.}
Obfuscation typically inflates code length and disrupts local regularities, resulting in extremely long sequences that exceed the context window of most open-source LLMs. Consequently, given an obfuscated snippet and assembly codes $c$, directly serializing it into a textual sequence for LLMs is impractical. To address this challenge, we employ a dedicated assembly encoder $F_{\mathrm{enc}}$ to compress obfuscated assembly into a compact representation with a fixed-dimensional space $\mathbb{R}^{d_{\text{enc}}}$, and further align this representation with the linguistic space of the LLM via an MLP-based projection $W : \mathbb{R}^{d_{\text{enc}}} \rightarrow \mathbb{R}^{d_{\text{llm}}}$, as shown in Eq.~\eqref{eq:assembly-embedding}, ensuring compatibility with limited context windows while preserving the essential structural and semantic information of the original assembly code.

\vspace{-4mm}
\begin{equation}
\small
\mathbf{Z}_{c} = W\bigl(F_{\mathrm{enc}}(c)\bigr), F_{\mathrm{enc}}(c) \in \mathbb{R}^{d_{\text{enc}}}, \;
\mathbf{Z}_{c} \in \mathbb{R}^{d_{\text{llm}}}
\label{eq:assembly-embedding}
\end{equation}
\vspace{-8mm}

\paragraph{Instruction Decoding with Special Tokens.}
After obtaining the compact assembly representation $\mathbf{Z}_{c}$, we condition the LLM on both encoded assembly and language instructions to generate analyzes. To explicitly structure the input, we introduce special tokens instead of treating assembly and instructions as homogeneous text. In particular, \texttt{<inst\_start>} and \texttt{<inst\_end>} delimit each instruction instance, separating prompt components and stabilizing instruction-conditioned decoding.

\vspace{-2mm}
\subsection{Phases I \& II: Supervised Knowledge Acquisition}
\label{sec:sft}
\vspace{-2mm}

In this subsection, we describe the training procedures for the projector and the instruction following capability of the LLM. The first two training phases are intentionally designed to be lightweight, providing a solid and efficient initialization for the subsequent self-evolution RL stage.

\vspace{-2mm}
\subsubsection{Phase I: Feature Space Alignment}
\label{sec:phase1}
\vspace{-2mm}
In the first phase, we freeze the LLM and only optimize the projector to align assembly features with the semantic space of the LLM. The projector is trained using the simplified summary $y$ generated by the frozen LLM from either the obfuscated or the original assembly representation $\mathbf{Z}_{c}$, as specified by the corresponding objective Eq.~\eqref{eq:l_align}. This ensures that the assembly token embedding acquires a direct and consistent semantic interpretation within the frozen LLM.

\vspace{-4mm}
\begin{equation}
\small 
\mathcal{L}_{\text{align}} = \mathbb{E}_{c \sim \mathcal{D}_{\text{align}}} \left[ - \sum_{i=1}^{L} \log P(y_i \mid \mathbf{Z}_{c}, y_{<i}) \right]
\label{eq:l_align}
\end{equation}
$\mathcal{D}_{\text{align}}$ is the input distribution used for alignment, and $L$ denotes the length of the output summary $y$.

\vspace{-2mm}
\subsubsection{Phase II: Instruction Fine-Tuning}
\label{sec:phase2}
\vspace{-2mm}
Subsequently, the projector and the LLM are jointly optimized on instruction following data while keeping the assembly encoder frozen. As formulated in Eq.~\eqref{eq:l_align}, this phase maintains the same autoregressive language-modeling loss as Phase I. This training process enables the LLM to internalize the semantics of obfuscated assembly and attain a foundational understanding of obfuscated assembly code, thereby establishing a robust precursor for the subsequent self-evolution stage.

\vspace{-2mm}
\subsection{Phase III: Self-Evolving RL with hybrid rewards}
\label{sec:htrtl}

\subsubsection{Self-Evolving RL}
Phases I--II establish robust general behaviors, commercial obfuscation remains challenging. Given the limited context length of open-source LLMs and the impracticality of collecting human annotations for long obfuscated-code analyses, we adopt a self-evolving RL scheme to enhance comprehension of obfuscated assembly. For each obfuscated assembly snippet $c$ and its corresponding instruction $q$, the model samples multiple candidate analyses, scores each candidate from multiple perspectives. These scores are then used to optimize the model itself.

\vspace{-2mm}
\subsubsection{Hybrid Rewards}

\label{sec:reward}
To provide reliable learning signals under heavily obfuscated assembly code, we redesign the reward function as a combination of structural reward and semantic reward.

\vspace{-2mm}
\paragraph{Structural Awareness Reward.}
The structural reward encourages the model to accurately identify obfuscation techniques while strictly penalizing the misclassification of obfuscated code as benign (ORG). Specifically, the reward is set to $1$ for a correct prediction of the specific obfuscation type, $0.2$ for the binary detection of whether obfuscation exists, and $0$ otherwise. Given the predicted class $\hat{t}$, the ground-truth $t_{\mathrm{gt}}$, and the set of obfuscation types $\mathcal{T}$, the reward $R_{\mathrm{struc}}$ is formally defined as:

\vspace{-4mm}
\begin{equation}
\small
R_{\mathrm{struc}} = 
\begin{cases}
1 & \text{if } \hat{t} = t_{\mathrm{gt}} \\
0.2 & \text{if } \hat{t} \neq t_{\mathrm{gt}} \text{ and } (\hat{t}, t_{\mathrm{gt}} \in \mathcal{T} \text{ or } \hat{t}, t_{\mathrm{gt}} \notin \mathcal{T}) \\
0 & \text{otherwise}
\end{cases}
\end{equation}
\vspace{-4mm}

\vspace{-2mm}
\paragraph{Semantic Reward.}
Relying solely on the correctness of the final output yields a sparse learning signal. To provide denser supervision, we introduce a semantic reward $R_{\mathrm{sem}}$ over the generated reasoning process. The model evaluates its own reasoning steps against the original unobfuscated assembly and the reference metadata along five dimensions, i.e., helpfulness, relevance, accuracy, detail, and comprehensiveness, from which $R_{\mathrm{sem}}$ is computed.

\vspace{-2mm}
\paragraph{Total Reward.}
The overall reward is a combination of structural and semantic components, weighted by a hyperparameter $\lambda$ and conditioned on the model's generated response $y$:

\vspace{-4mm}
\begin{equation}
\small
R_{\mathrm{total}} = \lambda R_{\mathrm{struc}}(\hat{t}, t_{\mathrm{gt}}) + (1-\lambda) R_{\mathrm{sem}}(c, q, y)
\end{equation}
\vspace{-8mm}

\subsubsection{Group Relative Policy Optimization}
\label{sec:grpo}
\vspace{-2mm}

We optimize our model using GRPO~\citep{shao2024deepseekmath}. For each sample, the model scores its own $K$ sampled trajectories to obtain rewards. We normalize rewards within each group and directly use the resulting standardized advantage in the GRPO objective:

\vspace{-4mm}
\begin{equation}
\small
\resizebox{0.92\hsize}{!}{$
\mathcal{J}_{\text{GRPO}}(\theta)
=
\mathbb{E}_{(c, q) \sim \mathcal{D}_{\text{rl}}}
\frac{1}{K}\sum_{k=1}^{K}
\left[
\operatorname{clip}\!\left(
\frac{\pi_\theta(y_k \mid c, q)}{\pi_{\theta_{\text{old}}}(y_k \mid c, q)}
\cdot
\frac{R_k - \mathrm{mean}(R)}{\mathrm{std}(R)},
\, 1-\epsilon,\, 1+\epsilon
\right)
-
\beta\, \mathbb{D}_{\text{KL}}(\pi_\theta \,\|\, \pi_{\text{ref}})
\right]
$}
\label{eq:grpo_obj}
\end{equation}

\vspace{-2mm}

$R_k$ is the reward of the $k$-th trajectory and $R=\{R_1, R_2, \dots, R_K\}$ denotes the set of trajectory rewards in the group. $\pi_\theta$ and $\pi_{\theta_{\text{old}}}$ are the new and old policies, and $\pi_{\text{ref}}$ is the reference policy.

\section{Experimental Setup}

\vspace{-2mm}
\subsection{Evaluation Benchmarks and Metrics}
\label{sec:eval_setup}
\vspace{-2mm}

\smallskip
\noindent
\textbf{Commercial Black-box VM-based Obfuscator Comprehension.}

We evaluate on \textbf{VMISA}~\citep{li2022chosen}, which tests whether models can recover the semantics of proprietary virtual instructions that are entirely unseen during training, making it a challenging out-of-distribution setting. It includes binaries protected by Code Virtualizer~\citep{CodeVirtualizerObf}, Themida (v3.0.7)~\citep{ThemidaObf}, and VMProtect (v3.5)~\citep{VMProtectObf}. We report the number of recovered instructions $k$ from the virtual instruction set (ISA) and the success rate $SR=\frac{k}{|\mathrm{ISA}|}\times 100\%$. Details are provided in  Appendix~\ref{appx:vmisa_setup_details}.

\vspace{-2mm}
\smallskip
\noindent
\textbf{Obfuscation-Aware Assembly-Language Instruction Following.}
We introduce \textbf{OASIF-Bench}, the benchmark dedicated to LLM-based obfuscated assembly comprehension and instruction following. We randomly select 30 domain-diverse assembly snippets from the BinaryCorp-3M test split, ensuring no overlap with the training data, while covering cryptography, malware, and protocol implementations. Following Sec.~\ref{sec:instruction_centric_data_generation}, we expand them into 150 snippets using five types (ORG, BCF, FLA, SUB and ALL). The resulting dataset contains 450 expert-curated queries spanning conversation, description, and reasoning tasks. All queries and reference answers are independently analyzed and cross-validated by five binary-analysis experts. Following the evaluation protocol in LLaVA~\citep{liu2024visual} and ASMA-Tune~\citep{wang2025asma}, GPT-5.4 serves as the evaluator and scores model outputs on a 1--10 scale across five dimensions (helpfulness, relevance, accuracy, detail, and comprehensiveness) relative to an expert-derived upper bound. Scores are normalized to a 0--100 scale for reporting, and each sample is scored over three independent trials to stabilize the estimate.

\vspace{-2mm}
\smallskip
\noindent
\textbf{Binary Code Similarity Detection (BCSD).}
For binary code similarity detection, we use seven established benchmarks widely adopted in prior BCSD research~\citep{su2024codeart, jiang2023nova, wang2025asma}: \textbf{Curl}, \textbf{Coreutils}, \textbf{Binutils}, \textbf{ImageMagick}, \textbf{SQLite}, \textbf{OpenSSL}, and \textbf{Putty}. Consistent with prior work, we report Recall@1 under varying pool sizes to measure the percentage of queries for which the ground-truth candidate is ranked first, and Mean Reciprocal Rank (MRR) as the average reciprocal rank of the correct candidate.

\vspace{-2mm}
\smallskip
\noindent
\textbf{Domain-Relevant Capability Preservation.}
To verify that OASIF preserves capabilities relevant to code generation and downstream binary-analysis tasks, we further evaluate on three non-overlapping benchmarks: HumanEval~\citep{chen2021evaluating}, VulBench~\citep{gao2023far}, and HumanEval-Decompile~\citep{tan2024llm4decompile}. We follow the official settings of each benchmark and report Pass@1 for HumanEval, Accuracy/F1/Precision/Recall for VulBench, and Compile Rate/Run Rate (Pass@1) for HumanEval-Decompile. Details of the dataset and optimization levels are provided in Appendix~\ref{appx:capability_benchmark_details}.

\vspace{-2mm}
\subsection{Baselines}
\label{sec:Baselines}
\vspace{-2mm}

We compare our framework against a representative suite of state-of-the-art LLMs:
(1) \textbf{Proprietary LLMs:} GPT-5, GPT-4-Turbo, Claude-Sonnet-4-5, Claude-Opus-4-6, and Gemini-2.5-Pro.
(2) \textbf{Open-Source Code LLMs:} Qwen2.5-Coder-Instruct (7B, 14B), Qwen3-Coder-Plus, DeepSeek-Coder-V2-Lite-Instruct-16B~\citep{zhu2024deepseek}, DeepSeek-Coder-Instruct-6.7B, GPT-oss-120B.

\vspace{-2mm}
\subsection{Model Instantiation and Training Settings}
\label{sec:model_training}
\vspace{-2mm}

We instantiate OASIF using three representative backbones: Qwen2.5-Coder-Instruct-14B, Qwen2.5-Coder-Instruct-7B, and DeepSeek-Coder-Instruct-6.7B.
The architecture comprises three components: an Assembly Code Encoder (CLAP-ASM~\citep{wang2024clap}, 110M parameters), a Connector (single-layer MLP, 30M parameters), and the LLM backbone. Dimension alignment adapts the projector output to each LLM's hidden size (e.g., 3,584 for Qwen2.5-7B and 4,096 for DeepSeek-6.7B).

Training is conducted in three stages.
Phase I (Alignment) is trained for 1 epoch on 8 NVIDIA H20 GPUs (learning rate $1\text{e-}5$, global batch size 1024).
Phase II (SFT) is fine-tuned for 1 epoch on 8 NVIDIA H20 GPUs (learning rate $1\text{e-}5$, global batch size 32).
Phase III (RL) is optimized for 3 epochs on 32 NVIDIA H20 GPUs, with 8 rollouts per instruction, a temperature of 0.7, top-$p$ 0.9, $\epsilon$=0.2, $\beta$=0.04, and a maximum sequence length of $L_{\max}$=4096. $\lambda$ is empirically set to 0.5 to balance the reward signals.

\vspace{-2mm}
\subsection{Implementation Details} 
\label{sec:Implementation Details}
\vspace{-2mm}
 
Inference is performed with a temperature of 0 and top-$k$ of 1 to ensure deterministic outputs. Prompts and data used for evaluation are strictly excluded from the retrieval corpus generation. BCSD follows the established few-shot CoT protocol from prior work~\citep{wang2025asma} to enable direct comparison of both general-purpose and specialized code LLMs. Under the deterministic decoding setting, small numerical differences are highly reproducible and are unlikely to be attributable to sampling variance.

\vspace{-2mm}
\section{Experiments}

\begin{table*}[t]
\vspace{-2mm}
\centering
\renewcommand{\arraystretch}{1.4}
\caption{Results on VMISA-Bench and OASIF-Bench. Best result per column among proprietary models and open-source models (including OASIF) are bolded.}

\makebox[\textwidth][r]{
\resizebox{1.005\textwidth}{!}{
{
\huge
\setlength{\tabcolsep}{2pt} 
\begin{tabular}{l l l l l l l l l l}
\toprule
\multirow{2}{*}{\textbf{Model}} 
& \multicolumn{3}{c}{\textbf{VMISA-Bench ($k$ (SR\%))}} 
& \multicolumn{6}{c}{\textbf{OASIF-Bench}} \\
\cmidrule(lr){2-4}\cmidrule(lr){5-10}
& \makecell{\textbf{Code} \\ \textbf{Virtualizer}} 
& \makecell{\textbf{Themida} \\ \textbf{v3.0.7}} 
& \makecell{\textbf{VMProtect} \\ \textbf{v3.5}} 
& \makecell{\textbf{ORG}} & \makecell{\textbf{BCF}} & \makecell{\textbf{FLA}} & \makecell{\textbf{SUB}} & \makecell{\textbf{ALL}} & \makecell{\textbf{Avg}} \\
\midrule
\multicolumn{10}{l}{\textit{\textcolor{gray}{Proprietary Large Language Models}}}\\
\midrule
GPT-5 
& 89 (64.5) 
& 45 (32.6) 
& 153 (64.6) 
& 76.8          & \textbf{79.1} & \textbf{80.0} & \textbf{77.0} & 74.0          & \textbf{77.4} \\
Claude-Sonnet-4-5 
& \textbf{104 (75.4)} 
& \textbf{57 (41.3)} 
& 105 (44.3) 
& 75.6          & 68.0          & 79.7          & 59.6          & 67.6          & 70.1 \\
Claude-Opus-4-6 
& 101 (73.2) 
& 56 (40.6) 
& 144 (60.8) 
& \textbf{80.2} & 76.9          & 78.5          & 67.2          & \textbf{74.8} & 75.5 \\
Gemini-2.5-Pro 
& 98 (71.0) 
& 41 (29.7) 
& \textbf{183 (77.2)} 
& 69.0          & 66.4          & 64.2          & 72.0          & 59.8          & 66.3 \\
\midrule
\multicolumn{10}{l}{\textit{\textcolor{gray}{Open-Source Code Large Language Models}}}\\
\midrule
Qwen2.5-Coder-Instruct-7B 
& 40 (29.0) 
& 5 (3.6) 
& 46 (19.4) 
& 37.8          & 35.8          & 41.8          & 37.7          & 41.8          & 39.0 \\
Qwen2.5-Coder-Instruct-14B 
& 60 (43.5) 
& 12 (8.7) 
& 78 (32.9) 
& 56.2          & 54.8          & 55.0          & 51.7          & 55.8          & 54.7 \\
Qwen3-Coder-Plus 
& 81 (58.7) 
& \textbf{23 (16.7)} 
& 74 (31.2) 
& 56.1          & 60.6          & \textbf{69.3} & 51.6          & 63.8          & 60.3 \\
DeepSeek-Coder-Instruct-6.7B 
& 32 (23.2) 
& 7 (5.1) 
& 52 (21.9) 
& 37.1          & 38.4          & 39.6          & 30.6          & 30.4          & 35.2 \\
DeepSeek-Coder-V2-Lite-Instruct-16B 
& 74 (53.6) 
& 22 (15.9) 
& 95 (40.1) 
& 58.0          & 52.0          & 56.0          & 50.0          & 51.0          & 53.4 \\
GPT-oss-120B 
& 42 (30.4) 
& 14 (10.1) 
& 92 (38.8) 
& \textbf{73.6} & \textbf{68.7} & 72.0          & 64.7          & \textbf{68.4} & \textbf{69.5} \\
\midrule
\multicolumn{10}{l}{\textit{\textbf{OASIF (Ours)}}}\\
\midrule
\rowcolor[gray]{0.95} \textbf{Qwen2.5-Coder-Instruct-7B-OASIF} 
& 46 (33.3)\srup{4.3} 
& 8 (5.8)\srup{2.2} 
& 56 (23.6)\srup{4.2} 
& 49.6\srup{11.8} & 38.6\srup{2.8}  & 44.4\srup{2.6}  & 46.4\srup{8.7}  & 49.8\srup{8.0}  & 45.8\srup{6.8} \\
\rowcolor[gray]{0.95} \textbf{Qwen2.5-Coder-Instruct-14B-OASIF} 
& \textbf{82 (59.4)\srup{15.9}} 
& 20 (14.5)\srup{5.8} 
& \textbf{118 (49.8)\srup{16.9}} 
& 62.0\srup{5.8}  & 61.7\srup{6.9}  & 68.4\srup{13.4} & \textbf{64.8\srup{13.1}} & 65.8\srup{10.0} & 64.5\srup{9.8} \\
\rowcolor[gray]{0.95} \textbf{DeepSeek-Coder-Instruct-6.7B-OASIF} 
& 52 (37.7)\srup{14.5} 
& 13 (9.4)\srup{4.3} 
& 82 (34.6)\srup{12.7} 
& 52.4\srup{15.3} & 35.0\srdown{3.4}
& 55.8\srup{16.2} & 40.9\srup{10.3} & 38.2\srup{7.8}  & 44.5\srup{9.3} \\
\bottomrule
\end{tabular}%
}}} 
\label{tab:vmisa_oasif_bench_unified}
\vspace{-6mm}
\end{table*}

\vspace{-2mm}
\subsection{VMISA-Bench Evaluation}\label{sec:vmisa_results}
\vspace{-2mm}

Tab.~\ref{tab:vmisa_oasif_bench_unified} reports the performance of OASIF and baseline models on VMISA-Bench, which constitutes a challenging out-of-distribution setting since the virtual ISAs are entirely unseen during training.

\vspace{-2mm}
\smallskip
\noindent
\textbf{OASIF consistently improves open-source LLMs across all three commercial VM obfuscators, with larger gains on smaller backbones, and remains effective under different VM instruction transformation regimes.}
Across Code Virtualizer, Themida, and VMProtect, OASIF-equipped models exhibit clear and stable gains over their corresponding backbones. For example, Qwen2.5-Coder-Instruct-14B improves from 43.5\% to 59.4\% SR on Code Virtualizer ($+15.9$), from 8.7\% to 14.5\% on Themida ($+5.8$), and from 32.9\% to 49.8\% on VMProtect ($+16.9$). Similar trends are observed for the 7B Qwen and DeepSeek-6.7B variants, indicating effectiveness across model scales and architectures, with diminishing returns as backbone capacity increases. The gains remain pronounced even on VMProtect, which combines one-to-many (O2M) and one-to-one (O2O) instruction mappings and semantically obfuscates context-switch instructions, requiring reasoning over multiple handler compositions for a single native instruction. Under this more complex regime, Qwen2.5-Coder-Instruct-14B-OASIF approaches or surpasses the success rates of some closed-source models on VMProtect despite using substantially fewer parameters in an open-source setting.

\vspace{-2mm}
\smallskip
\noindent
\textbf{These results indicate stronger obfuscation-aware comprehension rather than memorization of fixed mapping rules.}
By integrating token-efficient assembly representations with self-evolving reinforcement learning, OASIF improves generalization to unseen proprietary VM-based obfuscation and enhances compositional reasoning over virtual handlers in difficult OOD settings.

\vspace{-2mm}
\subsection{OASIF-Bench Evaluation}\label{sec:oasif_bench_results}
\vspace{-2mm}

Tab.~\ref{tab:vmisa_oasif_bench_unified} reports the performance of OASIF and baseline models on OASIF-Bench.

\vspace{-2mm}
\smallskip
\noindent
\textbf{OASIF consistently improves obfuscation-aware comprehension across all obfuscation types.}

Qwen2.5-Coder-Instruct-14B-OASIF improves the average score from 54.7 to 64.5 ($+9.8$), surpassing the roughly 30$\times$ larger Qwen3-Coder-Plus (60.3) and approaching several proprietary baselines. Qwen2.5-Coder-Instruct-7B-OASIF and DeepSeek-Coder-Instruct-6.7B-OASIF gain $+6.8$ and $+9.3$ on average, respectively, with positive improvements in nearly all obfuscation settings. These gains confirm that the improvements extend beyond VMISA-Bench alone and generalize to broader obfuscation-aware instruction-following settings. Length-stratified analysis refers to Appendix~\ref{appx:length_analysis}.

\begin{figure*}[t]
    \centering
    \vspace{-2mm}
    \includegraphics[width=1.0\linewidth]{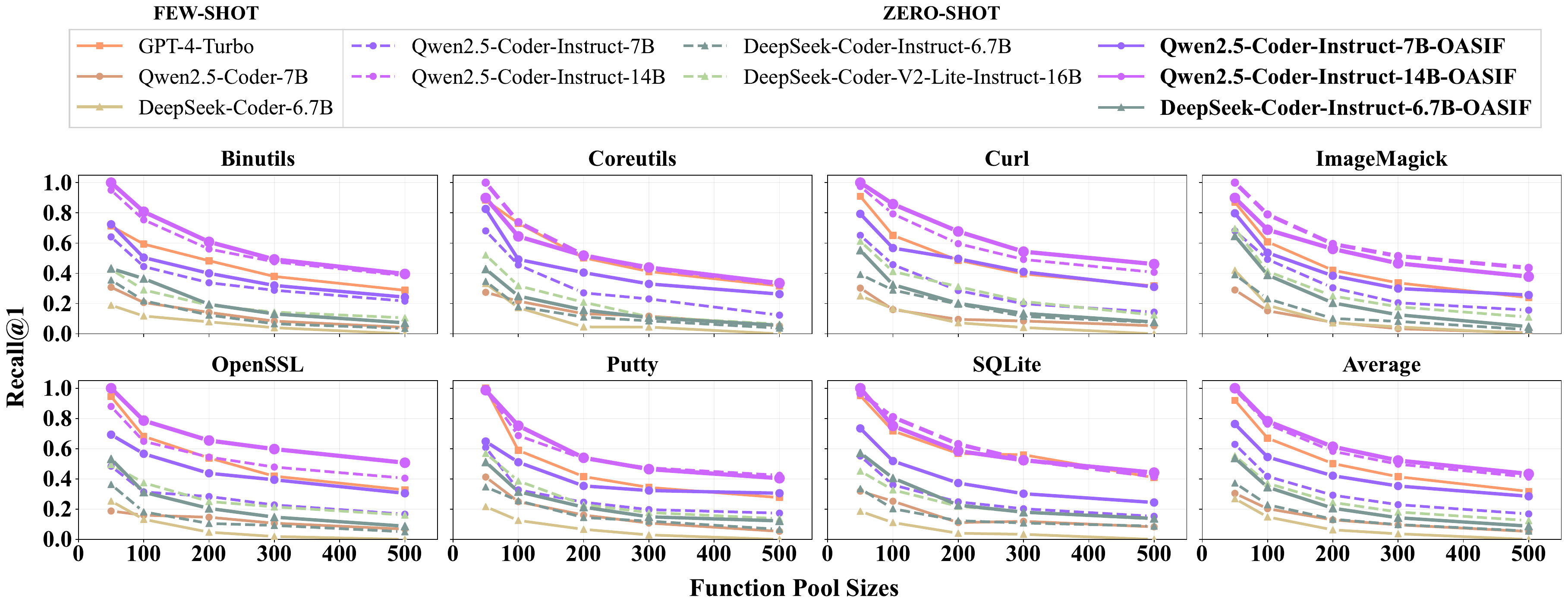}
    \vspace{-6mm}
    \caption{BCSD experimental results. We present normalized evaluation results. The x-axes denote different sizes of candidate function pools, and the y-axes denote the Recall@1 performance.}
    \label{fig:bcsd_recall1}
\end{figure*}

\begin{table*}[t]
\vspace{-2mm}
\centering
\small
\setlength{\tabcolsep}{4pt}
\renewcommand{\arraystretch}{1.15}
\caption{BCSD evaluation results across seven suites and the average score.}
\resizebox{\textwidth}{!}{
\begin{tabular}{l c c c c c c c c}
\toprule
\multirow{2}{*}{\textbf{Model}} &
\multicolumn{8}{c}{\textbf{BCSD Evaluation (Pool size 500)}} \\
\cmidrule(lr){2-9}
& \textbf{Binutils} & \textbf{Coreutils} & \textbf{Curl} & \textbf{ImageMagick} &
  \textbf{SQLite} & \textbf{OpenSSL} & \textbf{Putty} & \textbf{Average} \\
\midrule

\multicolumn{9}{l}{\textit{\textcolor{gray}{Few-shot CoT}}}\\
\midrule
GPT-4-Turbo
& 0.190 & 0.175 & 0.191 & 0.162 & 0.169 & 0.232 & 0.174 & 0.185 \\
Qwen2.5-Coder-7B
& 0.072 & 0.076 & 0.078 & 0.072 & 0.065 & 0.101 & 0.070 & 0.076 \\
DeepSeek-Coder-7B
& 0.050 & 0.053 & 0.052 & 0.074 & 0.037 & 0.067 & 0.047 & 0.054 \\
\midrule

\multicolumn{9}{l}{\textit{\textcolor{gray}{Zero-shot CoT}}}\\
\midrule
Qwen2.5-Coder-Instruct-7B
& 0.155 & 0.106 & 0.119 & 0.135 & 0.091 & 0.147 & 0.119 & 0.124 \\
Qwen2.5-Coder-Instruct-14B
& \textbf{0.227} & \textbf{0.179} & 0.224 & 0.226 & 0.172 & 0.246 & \textbf{0.212} & 0.212 \\
DeepSeek-Coder-Instruct-6.7B
& 0.075 & 0.070 & 0.083 & 0.084 & 0.066 & 0.095 & 0.075 & 0.078 \\
DeepSeek-Coder-V2-Lite-Instruct-16B
& 0.108 & 0.086 & 0.109 & 0.119 & 0.086 & 0.141 & 0.107 & 0.108 \\
\midrule

\multicolumn{9}{l}{\textit{\textbf{OASIF (Ours)}}}\\
\midrule
\rowcolor[gray]{0.95} \textbf{Qwen2.5-Coder-Instruct-7B-OASIF}
& 0.166 & 0.149 & 0.176 & 0.198 & 0.118 & 0.208 & 0.159 & 0.168 \\
\rowcolor[gray]{0.95} \textbf{Qwen2.5-Coder-Instruct-14B-OASIF}
& 0.225 & 0.174 & \textbf{0.236} & \textbf{0.247} & \textbf{0.175} & \textbf{0.265} & 0.208 & \textbf{0.218} \\
\rowcolor[gray]{0.95} \textbf{DeepSeek-Coder-Instruct-6.7B-OASIF}
& 0.082 & 0.077 & 0.080 & 0.099 & 0.086 & 0.110 & 0.097 & 0.090 \\
\bottomrule
\end{tabular}
}
\vspace{-6mm}

\label{tab:bcsd_eval}

\end{table*}

\vspace{-4mm}
\subsection{Binary Code Similarity Detection Evaluation}\label{sec:main_result_3_bcsd}
\vspace{-2mm}

Fig.~\ref{fig:bcsd_recall1} reports Recall@1 under varying pool sizes, while Tab.~\ref{tab:bcsd_eval} presents Mean Reciprocal Rank (MRR) results with a fixed pool size of 500 across seven suites.

\vspace{-2mm}
\smallskip
\noindent
\textbf{OASIF consistently improves binary code similarity detection across model backbones and suites, with larger gains on weaker backbones and diminishing returns on stronger ones.}

As shown in Fig.~\ref{fig:bcsd_recall1}, OASIF-enhanced models maintain higher average Recall@1 than their baselines across pool sizes. The same trend appears in Tab.~\ref{tab:bcsd_eval}, Qwen2.5-Coder-Instruct-7B-OASIF improves average MRR by $+0.044$ with gains on all seven benchmarks, DeepSeek-Coder-Instruct-6.7B-OASIF improves by $+0.012$, and Qwen2.5-Coder-Instruct-14B-OASIF still gains $+0.006$ despite the stronger baseline. Notably, the 14B backbone already exceeds GPT-4-Turbo with few-shot CoT before OASIF training (0.212 vs.~0.185), indicating limited headroom on this retrieval protocol. Overall, the larger gains on the 7B and DeepSeek-6.7B backbones suggest that OASIF primarily improves assembly comprehension, with diminishing returns on already strong backbones.

\begin{table*}[t]
\vspace{-2mm}
\centering
\setlength{\tabcolsep}{2pt}
\renewcommand{\arraystretch}{1.15}
\caption{Capability preservation across HumanEval, VulBench, and HumanEval-Decompile.}
\resizebox{\textwidth}{!}{
\begin{tabular}{l c cccc cc}
\toprule
\multirow{2}{*}{\textbf{Model}}
& \multicolumn{1}{c}{\textbf{HumanEval}}
& \multicolumn{4}{c}{\textbf{VulBench}}
& \multicolumn{2}{c}{\textbf{HumanEval-Decompile}} \\
\cmidrule(r{2pt}){2-2}\cmidrule(l{2pt}r{2pt}){3-6}\cmidrule(l{2pt}){7-8}
& \textbf{Pass@1 (\%)}
& \textbf{Acc (\%)} & \textbf{F1 (\%)} & \textbf{Prec (\%)} & \textbf{Recall (\%)}
& \textbf{Compile Rate (\%)} & \textbf{Run Rate (\%)} \\
\midrule
\multicolumn{8}{l}{\textit{\textcolor{gray}{Open-Source Code Large Language Models}}}\\
\midrule
Qwen2.5-Coder-Instruct-7B           & 88.4           & 41.28          & 29.20          & 27.31          & 51.67          & 81.25          & 4.27 \\
Qwen2.5-Coder-Instruct-14B          & 89.6           & 59.13          & 33.46          & 33.69          & 45.28          & 80.03          & 6.86 \\
DeepSeek-Coder-Instruct-6.7B        & 66.1           & 32.57          & 26.42          & 24.02          & 52.84          & 48.48          & 3.36 \\
DeepSeek-Coder-V2-Lite-Instruct-16B & 81.1           & 48.02          & 27.08          & 24.01          & 49.33          & 53.66          & 3.81 \\
\midrule
\multicolumn{8}{l}{\textit{\textbf{OASIF (Ours)}}}\\
\midrule
\rowcolor[gray]{0.95} \textbf{Qwen2.5-Coder-Instruct-7B-OASIF}    & 86.5           & 40.93          & 27.57          & 27.04          & 48.02          & 83.54          & 3.81          \\
\rowcolor[gray]{0.95} \textbf{Qwen2.5-Coder-Instruct-14B-OASIF}   & 84.8           & 58.60          & 32.20          & 33.30          & 44.57          & 82.10          & 6.20 \\
\rowcolor[gray]{0.95} \textbf{DeepSeek-Coder-Instruct-6.7B-OASIF} & 50.6           & 31.73          & 26.03          & 23.44          & 52.78          & 50.15          & 2.44          \\
\bottomrule
\end{tabular}
}
\label{tab:vulbench_decompile}
\vspace{-4mm}
\end{table*}
\begin{table}[h]
\centering
\begin{minipage}[h]{0.44\textwidth}
\begin{minipage}[t]{\linewidth}
\centering
\small
\setlength{\tabcolsep}{12pt}
\captionof{table}{Ablation study on different configurations and training components (Recall@1 on the average BCSD evaluation sets).}
\resizebox{\linewidth}{!}{
\begin{tabular}{l l}
\toprule
\multicolumn{2}{l}{\textbf{Training Components}} \\
\midrule
\textbf{Configuration} & \textbf{Recall@1 (\%)} \\
\midrule
\rowcolor[gray]{0.95} Full training & \textbf{49.1} \\
w/o Assembly encoder & 41.7 \, \textcolor{gray}{($\downarrow 7.4$)} \\
w/o Feature Space Alignment & 27.3 \, \textcolor{gray}{($\downarrow 21.8$)} \\
w/o Instruction Fine-tuning & 23.6 \, \textcolor{gray}{($\downarrow 25.5$)} \\
w/o Self-Evolving RL & 39.7 \, \textcolor{gray}{($\downarrow 9.4$)} \\
\midrule
\multicolumn{2}{l}{\textbf{Model Scale}} \\
\midrule
7B model size & 35.9 \, \textcolor{gray}{($\downarrow 13.2$)} \\
14B model size & 49.1 \\
\bottomrule
\end{tabular}
}
\label{tab:ablation_recall1}
\end{minipage}%
\end{minipage}
\hfill
\begin{minipage}[h]{0.51\textwidth}
\begin{minipage}[t]{\linewidth}
\centering
\small
\setlength{\tabcolsep}{3pt}
\renewcommand{\arraystretch}{1.15}
\captionof{table}{Two-stage agentic loop evaluation on VMProtect v3.5.}
\resizebox{\linewidth}{!}{
\begin{tabular}{l cc}
\toprule
\textbf{Configuration} & \textbf{$k$} & \textbf{SR (\%)} \\
\midrule
\multicolumn{3}{l}{\textit{\textcolor{gray}{Direct inference}}} \\
\midrule
Claude-Opus-4-6                        & 144 & 60.8 \\
Qwen3-Coder-Plus                       & 74  & 31.2 \\
GPT-oss-120B                           & 92  & 38.8 \\
Qwen2.5-Coder-Instruct-14B-OASIF       & 118 & 49.8 \\
\midrule
\multicolumn{3}{l}{\textit{\textcolor{gray}{Two-stage agentic loop (back-end: Claude-Opus-4-6)}}} \\
\midrule
\quad Front-end: Qwen3-Coder-Plus                              & 120 & 50.6 \\
\quad Front-end: GPT-oss-120B                                  & 147 & 62.0 \\
\rowcolor[gray]{0.95} \quad Front-end: \textbf{Qwen2.5-Coder-Instruct-14B-OASIF} & \textbf{148} & \textbf{62.4} \\
\bottomrule
\end{tabular}
}
\label{tab:agentic_loop}
\end{minipage}
\end{minipage}
\vspace{-4mm}
\end{table}

\vspace{-2mm}
\subsection{Capability Preservation}\label{sec:capability_preservation}
\vspace{-2mm}
To assess whether OASIF training preserves general and domain-relevant capabilities, we evaluate on three benchmarks that lie outside the OASIF training distribution. Results are presented in Tab.~\ref{tab:vulbench_decompile}.

\vspace{-2mm}
\smallskip
\noindent
\textbf{OASIF preserves domain-relevant capabilities while keeping general capability trade-offs within an acceptable range.}

HumanEval Pass@1 decreases moderately for Qwen2.5-Coder-Instruct-7B ($-1.9$) and 14B ($-4.8$), with a larger drop on DeepSeek-Coder-Instruct-6.7B. In contrast, VulBench remains stable, with all four metrics within roughly one point of the base models, and HumanEval-Decompile shows consistent Compile Rate gains of $+2.29$, $+2.07$, and $+1.67$ across the three backbones, suggesting mild positive transfer to decompilation comprehension. Overall, OASIF improves obfuscation comprehension while largely preserving domain-relevant performance.

\vspace{-2mm}
\subsection{Ablation Study}\label{sec:ablation}
\vspace{-2mm}

Tab.~\ref{tab:ablation_recall1} reports the ablation results of key components in the proposed framework.

\vspace{-2mm}
\smallskip
\noindent
\textbf{The effectiveness of the framework relies on the coordinated design of representation, alignment, and optimization stages.}
Removing any major component consistently degrades performance from the full configuration (49.1\%), indicating that the observed gains do not arise from a single module. Feature space alignment and instruction fine-tuning play a central role in bridging assembly representations and language-model reasoning. The performance decrease observed when removing the self-evolving RL further confirms that iterative behavior refinement provides additional gains.
The 14B backbone outperforms the 7B, demonstrating that larger models are better able to leverage the aligned representations and training signals introduced by the framework.
The ablation results validate the necessity of each component within the unified training pipeline.

\vspace{-2mm}
\subsection{Agentic Pipeline Evaluation}\label{sec:agentic_pipeline}
\vspace{-2mm}

We study a two-stage agentic pipeline on VMProtect v3.5: a front-end model analyses the obfuscated binary, and a fixed back-end model performs final semantic reasoning based on the observations and original assembly. This setup isolates the effect of front-end comprehension (Tab.~\ref{tab:agentic_loop}).

\vspace{-2mm}
\smallskip
\noindent
\textbf{Improving front-end comprehension strengthens the full pipeline.}
Qwen3-Coder-Plus as a front-end still underperforms compared to standalone Claude-Opus-4-6, suggesting an observation bottleneck. GPT-oss-120B raises SR to 62.0\%, and the OASIF-trained 14B front-end further reaches 62.4\% at $k{=}148$, showing that OASIF improves observation quality and enables a 14B model to effectively augment a frontier reasoning model.
\vspace{-2mm}
\section{Conclusion}
\vspace{-2mm}
We present OASIF,  an Obfuscation-Aware Self-evolving Instruction-Following framework with three-stage training for obfuscated assembly instruction following and comprehension. Across commercial VM-based obfuscation, OASIF-Bench, seven BCSD benchmarks, OASIF consistently improves open-source LLMs while preserving capabilities on HumanEval, VulBench, and HumanEval-Decompile.

\vspace{-2mm}
\section*{Limitations}
\label{sec:limitations}

While OASIF delivers consistent improvements in obfuscated assembly comprehension, several limitations remain and point to directions for future work.

\vspace{-2mm}
\paragraph{Granularity and scope of analysis.}
OASIF operates at the function-level on x86\_64 binaries, following the standard workflow in which disassemblers (e.g., IDA Pro, Ghidra) decompose binaries into individual functions. End-to-end analysis of complete large-scale binaries with inter-procedural reasoning, cross-function data-flow tracking, and whole-program symbolic recovery is beyond the current scope and constitutes an important direction for future work. Generalization to other instruction set architectures (e.g., ARM, RISC-V) has also not been empirically validated.

\vspace{-2mm}
\paragraph{Statistical reporting under high compute cost.}
Each OASIF training run requires 8~H20 GPUs for Phases I--II and 32~H20 GPUs for Phase III RL, making multi-seed sweeps prohibitive within our budget. We therefore report single-run results under near-deterministic inference (temperature $=0$, top-$k=1$), which yield highly reproducible outputs. The consistency of the observed gains across three different backbones (Qwen2.5-Coder-Instruct-7B/14B, DeepSeek-Coder-Instruct-6.7B) and across multiple independent benchmarks (VMISA-Bench, OASIF-Bench, BCSD, HumanEval, VulBench, HumanEval-Decompile) mitigates but does not fully replace formal variance estimation.

\vspace{-2mm}
\paragraph{Reliance on synthetic supervision and expert curation.}
The three-phase training data are generated by an LLM $M_\phi$ with few-shot prompts and are subsequently filtered. $D_{\text{sft}}$ and especially $D_{\text{rl}}$ require manual expert verification, which caps dataset size (4{,}244 RL samples in our setting) and introduces annotator-specific biases. Scaling to substantially larger RL corpora while maintaining quality remains an open problem, and the semantic reward $R_{\text{sem}}$ obtained from LLM self-evaluation may still contain residual judge biases that our decoupled structural reward only partially offsets.

\vspace{-2mm}
\paragraph{Coverage of obfuscation schemes.}
Training supervision comes from OLLVM-based variants (SUB, BCF, FLA), whereas evaluation targets three commercial black-box VM obfuscators (Code Virtualizer, Themida~v3.0.7, VMProtect~v3.5). Although this represents a strong OOD evaluation, the space of real-world obfuscation is broader (packers, hardware-assisted protections, anti-analysis techniques, novel or evolving commercial protectors), and robustness to obfuscators whose design departs substantially from the VM-based family we studied is not guaranteed.

\vspace{-2mm}
\paragraph{Backbone-dependent preservation of general capability.}
The impact of OASIF training on unrelated general code generation depends on the backbone. Qwen2.5-Coder backbones exhibit only modest HumanEval drops, while DeepSeek-Coder-Instruct-6.7B shows a larger decline that is not observed on binary-oriented benchmarks (VulBench, HumanEval-Decompile). This suggests that the trade-off between obfuscation-specialized reasoning and broad code generation is model-dependent, and principled techniques for capability preservation (e.g., adapter-based training, replay buffers of general code) are left for future work.

\vspace{-2mm}
\paragraph{Dual-use considerations.}
OASIF is designed to advance defensive reverse engineering, malware analysis, and vulnerability triage; however, the same capability can, in principle, assist actors seeking to analyze protected software or adapt obfuscated malware. We restrict releases to research benchmarks and model weights for reproducibility, avoid releasing ready-to-use deobfuscation pipelines, and emphasize that deployment should comply with applicable laws, licenses, and responsible-disclosure practices. A fuller discussion is provided in the Broader Impact Statement.


\section*{Broader Impact Statement}

This paper advances machine learning methods for automated understanding of obfuscated assembly code, with potential benefits for software security, including improved reverse engineering, malware analysis, vulnerability triage, and defensive deobfuscation. At the same time, the techniques may be misused to accelerate analysis of protected software, weaken intellectual property protections, or aid malicious actors in understanding and adapting obfuscated malware. To mitigate these risks, we focus on evaluation in research benchmarks, avoid releasing actionable deobfuscation tooling or exploitation guidance, and emphasize that deployment should follow applicable laws, licenses, and responsible disclosure practices. We hope this work encourages further research on robust, auditable, and security-oriented LLMs, alongside community development of appropriate safeguards and usage policies.


\bibliography{main}
\bibliographystyle{plainnat}


\newpage
\appendix
\section{Dataset Composition}
\label{appx:dataset_composition}

We report the full composition of the raw assembly corpus, the OLLVM variant breakdown on Juliet (x86\_64), and the three-stage training datasets in Tab.~\ref{tab:datasets_description}. All variant-wise counts and the exact OLLVM compilation flags are also included for reproducibility.

\begin{table}[h]
  \centering
  \caption{Dataset composition (raw corpus, OLLVM variants, and three-stage training).}
  \vspace{2mm}
  \label{tab:datasets_description}
  \setlength{\tabcolsep}{5pt}
  \renewcommand{\arraystretch}{1.15}
  \small
  \begin{threeparttable}
  \begin{tabularx}{0.65\linewidth}{@{}X l@{}}
    \toprule
    \textbf{Dataset} & \textbf{Records} \\
    \midrule

    \multicolumn{2}{@{}l}{\textbf{Raw Assembly Corpus}} \\
    \midrule
    BinaryCorp-3M (train)                    & 212{,}117 \\
    Juliet (unobfuscated)                    & 79{,}920 \\
    Juliet (OLLVM-obfuscated)\tnote{a}       & 221{,}069 \\
    \midrule

    \multicolumn{2}{@{}l}{\textbf{OLLVM Variant Breakdown (Juliet, x86\_64)}} \\
    \midrule
    None (no obfuscation)                    & 79{,}920 \\
    SUB (Instruction Substitution)\tnote{b}  & 2{,}651  \\
    FLA (Control-Flow Flattening)\tnote{c}   & 58{,}578 \\
    BCF (Bogus Control Flow)\tnote{d}        & 79{,}920 \\
    ALL (SUB+FLA+BCF)\tnote{e}               & 79{,}920 \\
    \midrule

    \multicolumn{2}{@{}l}{\textbf{Training Data}} \\
    \midrule
    $\mathcal{D}_{\text{align}}$ (simp) & 430{,}027 \\
    $\mathcal{D}_{\text{sft}}$ (detail/conv/reason) & 11{,}750 \\
    $\mathcal{D}_{\text{rl}}$ (reason$^{+}$) & 4{,}244 \\
    \bottomrule
  \end{tabularx}

\begin{tablenotes}[flushleft]
\scriptsize
\setlength{\tabcolsep}{3pt}
\begin{tabularx}{\linewidth}{@{}l l >{\raggedright\arraybackslash}X@{}}
$^{\text{a}}$ & Include & \nolinkurl{sub/fla/bcf/all} \\
$^{\text{b}}$ & SUB & \nolinkurl{-mllvm -sub} \\
$^{\text{c}}$ & FLA & \nolinkurl{-mllvm -fla -mllvm -perFLA=100} \\
$^{\text{d}}$ & BCF & \nolinkurl{-mllvm -bcf -mllvm -boguscf-prob=100 -mllvm -boguscf-loop=1} \\
$^{\text{e}}$ & ALL & \nolinkurl{-mllvm -sub -mllvm -fla -mllvm -perFLA=100 -mllvm -bcf -mllvm -boguscf-prob=100 -mllvm -boguscf-loop=1} \\
\end{tabularx}
\item[]
\end{tablenotes}

  \end{threeparttable}
  
\end{table}

\section{Training Data Construction and Quality Control}
\label{appx:data_quality_control}

We organize the training data into three stage-specific subsets: $D_{\text{align}}$, $D_{\text{sft}}$, and $D_{\text{rl}}$. Beyond the aggregate sizes reported in Sec.~\ref{sec:dataset_engine}, we summarize the stage-wise source composition and filtering criteria here to clarify data quality control.

For $D_{\text{align}}$, we apply automated deduplication by function name and description to remove exact and near-duplicate samples. Starting from the 212{,}117 BinaryCorp-3M, 79{,}920 unobfuscated Juliet, and 221{,}069 OLLVM-obfuscated Juliet snippets reported in Tab.~\ref{tab:datasets_description}, this filtering yields the final 430{,}027 aligned instances used in Phase~I.

For $D_{\text{sft}}$, we retain only samples whose obfuscated assembly is at least 1.5$\times$ longer than the corresponding original assembly, ensuring that the transformation meaningfully changes the code structure. We further apply Spark TopK filtering with $K{=}80$ to balance token-length distributions across obfuscation types. After these automated filters, three domain experts independently review the retained samples to verify that the generated question--answer pairs are factually accurate and that the answers correctly describe the behavior of the obfuscated code; any sample flagged by at least one reviewer is removed. The final split contains about 12K instances, including 2{,}684 non-obfuscated samples and 9{,}066 obfuscated samples (SUB: 186, BCF: 3{,}252, FLA: 3{,}234, ALL: 2{,}394).

For $D_{\text{rl}}$, all retained question--answer pairs require multi-step reasoning and complete reasoning traces. We apply manual filtering, deduplication, and expert verification, and all candidate samples are independently reviewed by three binary-analysis experts. Each expert evaluates whether the question is meaningful and whether the reference answer is correct. We retain only samples that receive unanimous agreement from all three experts. The final split contains 4{,}244 instances, with 1{,}058 ORG, 30 SUB, 1{,}054 BCF, 1{,}048 FLA, and 1{,}054 ALL samples.

\section{Detailed VMISA Benchmark Description}
\label{appx:vmisa_setup_details}

VMISA~\citep{li2022chosen} contains programs protected by commercial black-box virtualization obfuscators, including Code Virtualizer~\citep{CodeVirtualizerObf}, Themida (v3.0.7)~\citep{ThemidaObf}, and VMProtect (v3.5)~\citep{VMProtectObf}, and tests whether models can recover the semantics of proprietary virtual instructions unseen during training, providing a strong out-of-distribution evaluation. Native-instruction to virtual-handler mappings are categorized into two families: One-to-One (O2O), where a single native instruction is emulated by one handler typically containing a kernel operation aligned with the original mnemonic, and One-to-Multiple (O2M), where one native instruction is realized by a composition of multiple handlers, potentially with multiple mapping rules for the same operation (e.g., implementing \texttt{xor} via logical \texttt{NOR}/\texttt{NAND} handlers rather than a dedicated \texttt{xor} handler). In practice, Code Virtualizer and Themida adopt O2O mappings over a virtual ISA of size $|\mathrm{ISA}|{=}138$, whereas VMProtect combines O2M and O2O mappings over $|\mathrm{ISA}|{=}237$, yielding a more diverse mapping space.

The benchmark also stresses analysis under context-switch obfuscation. While VMs typically save and restore native context using \texttt{mov} or \texttt{push}/\texttt{pop}, Code Virtualizer and Themida additionally use \texttt{xchg} and insert decoy context switches that may occur at the same stack depth as true restorations, complicating context pairing. VMProtect instead semantically obfuscates the generated context-switch instructions, hindering rule-based simplification.

For evaluation, we extract the virtual instruction set (ISA) of each obfuscator and report the number of recovered instructions $k$, where a virtual instruction is considered recovered if the model correctly identifies the corresponding Intel-ISA semantics of a virtual handler against expert-annotated ground-truth mappings. Recovery requires the correct native mnemonic and operand semantics; for O2M transformations in VMProtect, it further requires identifying the complete semantic equivalence of the handler composition. We report the success rate $SR=\frac{k}{|\mathrm{ISA}|}\times 100\%$ to normalize across different ISA sizes.

\section{Detailed Capability-Preservation Benchmark Description}
\label{appx:capability_benchmark_details}

To verify that OASIF preserves capabilities relevant to code generation and downstream binary-analysis tasks, we evaluate on three additional benchmarks that do not overlap with the OASIF training data. \textbf{HumanEval} comprises 164 hand-written Python programming tasks with function signatures and docstrings, and is widely used to assess code LLMs~\citep{guo2024deepseek, hui2024qwen2}. \textbf{VulBench}~\citep{gao2023far} targets vulnerability detection and covers three task types and seven sub-datasets (big-vul, ctf\_ida, ctf\_reversed, d2a, devign, magma\_bin, magma\_src), totaling 9{,}222 test cases. \textbf{HumanEval-Decompile}~\citep{tan2024llm4decompile} is a binary-to-source decompilation benchmark comprising 164 C functions aligned with HumanEval and compiled at optimization levels O0--O3.

We follow the official settings of each benchmark. Specifically, for HumanEval we adopt Pass@1 to measure the proportion of functionally correct code generations that pass all test cases. For VulBench, we report Accuracy, F1, Precision, and Recall aggregated across all task types. For HumanEval-Decompile, we report Compile Rate and Run Rate (Pass@1) averaged across optimization levels O0--O3.

\section{Length-Stratified Evaluation on OASIF-Bench}
\label{appx:length_analysis}

To better understand how OASIF behaves across input scales, we partition OASIF-Bench samples into three token-length buckets and evaluate Qwen2.5-Coder-Instruct-7B (\textit{Base}) and its OASIF-trained counterpart (\textit{OASIF}) on each bucket. Per-dimension scores are reported on a 0--10 scale; the Overall column is the per-bucket average across the five dimensions, and Weighted Avg is aggregated over all buckets. Results are reported in Tab.~\ref{tab:length_stratified}.

Both the base and OASIF-trained models exhibit a clear inverse relationship between input length and output quality, and the gain introduced by OASIF is most pronounced in the short-to-medium bucket (291--4{,}719 tokens: +0.67 Overall) while remaining positive in the longest bucket (10{,}391--46{,}567 tokens: +0.58 Overall). Since the assembly encoder compresses arbitrary-length assembly into a single fixed-dimensional embedding mapped to one LLM token, longer inputs inevitably lose fine-grained details, which explains the steeper degradation of Relevance. Long sequences (10K+) also fall in the tail of the encoder's pretraining distribution, further reducing encoding quality.

\begin{table}[h]
\centering
\small
\setlength{\tabcolsep}{4pt}
\renewcommand{\arraystretch}{1.15}
\caption{Length-stratified evaluation on OASIF-Bench.}
\vspace{2mm}
\resizebox{1\textwidth}{!}{
\begin{tabular}{l l cccccc}
\toprule
\textbf{Model} & \textbf{Token Range} & \textbf{Helpfulness} & \textbf{Relevance} & \textbf{Accuracy} & \textbf{Detail} & \textbf{Comprehensiveness} & \textbf{Overall} \\
\midrule
\multirow{4}{*}{\shortstack[l]{Qwen2.5-Coder-\\Instruct-7B\\(Base)}}
& 291~--~4{,}719     & 3.85 & 6.50 & 3.60 & 3.25 & 3.20 & 4.08 \\
& 5{,}226~--~9{,}648 & 3.40 & 5.55 & 3.50 & 2.78 & 3.07 & 3.66 \\
& 10{,}391~--~46{,}567 & 3.35 & 5.20 & 3.25 & 2.90 & 2.90 & 3.52 \\
& \textbf{Weighted Avg} & \textbf{3.69} & \textbf{6.21} & \textbf{3.53} & \textbf{3.13} & \textbf{3.13} & \textbf{3.90} \\
\midrule
\multirow{4}{*}{\shortstack[l]{Qwen2.5-Coder-\\Instruct-7B-\\OASIF (Ours)}}
& 291~--~4{,}719     & 4.66 & 7.44 & 4.41 & 3.57 & 3.66 & 4.75 \\
& 5{,}226~--~9{,}648 & 4.30 & 6.43 & 5.55 & 3.05 & 3.17 & 4.50 \\
& 10{,}391~--~46{,}567 & 4.20 & 5.80 & 3.90 & 3.30 & 3.30 & 4.10 \\
& \textbf{Weighted Avg} & \textbf{4.52} & \textbf{7.07} & \textbf{4.41} & \textbf{3.44} & \textbf{3.52} & \textbf{4.58} \\
\bottomrule
\end{tabular}
}
\label{tab:length_stratified}
\vspace{-4mm}
\end{table}

\section{Reward Design and Evaluation Decoupling}
\label{appx:reward_evaluation_details}

The semantic reward $R_{\text{sem}}$ is computed by a privileged reward model (PRM) that conditions on the original source code $c_{\text{src}}$ and obfuscation parameters $\Omega$, neither of which is available to the policy model $\pi_{\theta}$. This information asymmetry is intentional: the PRM evaluates semantic faithfulness with privileged posterior information rather than scoring outputs based only on the surface plausibility of the model response. As a result, the reward process is structurally decoupled from policy generation and does not reduce to self-evaluation.

We further decouple training-time reward computation from benchmark-time scoring at three levels. First, during RL, the hybrid reward combines structural and semantic signals, with $\lambda{=}0.5$ used as an empirical balancing coefficient between the two terms. Second, OASIF-Bench is evaluated by GPT-5.4 as an independent external judge against expert-derived reference answers across five dimensions: helpfulness, relevance, accuracy, detail, and comprehensiveness. Third, VMISA-Bench, BCSD, VulBench, and HumanEval-Decompile rely on external benchmarks and established evaluation protocols that are independent of the reward model. The consistency of the observed gains across these heterogeneous settings provides cross-validation beyond any single evaluator.

\section{Notes on BCSD Evaluation Protocol}
\label{appx:bcsd_protocol}

BCSD is included as a capability-preservation benchmark rather than as the primary measure of obfuscation-aware instruction following. Following prior work, all models generate natural-language descriptions that are then compared through semantic retrieval using Contriever-MSMARCO. We report Recall@1 under varying pool sizes and Mean Reciprocal Rank (MRR) with a fixed pool size of 500.

To enable direct comparison with published BCSD results from both general-purpose and specialized code LLMs, we follow the established few-shot chain-of-thought protocol used in prior work. Inference is performed under near-deterministic settings (temperature 0 and top-$k$ 1), and prompts and evaluation data are strictly excluded from the retrieval-corpus construction process. Under this protocol, small numerical differences are highly reproducible and are unlikely to be explained by sampling variance alone.

\section{Practical Scope on Complex Software}
\label{appx:whole_binary_scope}

VMISA-Bench already evaluates real commercial obfuscators applied to actual programs. For whole-binary analysis of complex software, OASIF is intended to operate within the standard reverse-engineering workflow. Tools such as IDA Pro or Ghidra first decompose the binary into individual functions, after which OASIF is applied at the function level. This function-level granularity is consistent with how human analysts and existing automated systems analyze real-world binaries.

The broader applicability is also supported by the source distribution of OASIF-Bench, whose assembly snippets are drawn from the BinaryCorp-3M test split, spanning diverse real software domains, and by the length-stratified results in Appendix~\ref{appx:length_analysis}, which show consistent gains across inputs ranging from hundreds to more than 40K tokens. End-to-end inter-procedural reasoning over entire large-scale binaries remains outside the current scope and is left for future work.



\end{document}